\documentclass[12pt]{article} 
\usepackage{graphicx}
\newcommand{\be}{\begin{equation}}
\newcommand{\ee}{\end{equation}}
\newcommand{\bear}{\begin{eqnarray}}
\newcommand{\eear}{\end{eqnarray}}
\newcommand{\ba}{\begin{array}}
\newcommand{\ea}{\end{array}}

\begin{document}

\begin{titlepage}
\vfill
\begin{flushright}
{\normalsize IC/2009/029}\\
{\normalsize arXiv:xxx.xxxx [hep-th]}\\
\end{flushright}

\vfill
\begin{center}
{\Large\bf Parity asymmetric boost invariant plasma  in AdS/CFT correspondence }

\vskip 0.3in
{Ho-Ung  Yee\footnote{\tt hyee@ictp.it}}

\vskip 0.15in

 {\it ICTP, High Energy, Cosmology and Astroparticle Physics,} \\
{\it Strada Costiera 11, 34014, Trieste, Italy}
\\[0.3in]

{\normalsize  2009}

\end{center}

\vfill

\begin{abstract}

We consider a simple extension to the previously found gravity solution corresponding to a boost invariant Bjorken plasma, by allowing components that are asymmetric under parity flipping of the spacetime rapidity.
Besides the question whether this may have a realization in collisions of different species of projectiles, such as lead-gold collision, our new time-dependent gravity background can serve as a test ground
for the recently proposed second order conformal viscous hydrodynamics.
We find that non-trivial parity-asymmetric effects start to appear at second order
in late time expansion, and we map the corresponding energy-momentum tensor to the
second order conformal hydrodynamics to find certain second order transport coefficients.
Our results are in agreement with the previous results in literature, giving one more corroborative
evidence for the validity of the framework.

\end{abstract}

\vfill

\end{titlepage}
\setcounter{footnote}{0}

\baselineskip 18pt \pagebreak
\renewcommand{\thepage}{\arabic{page}}
\pagebreak

\section{Introduction and motivation}

Since the advent of the RHIC experiment, dynamics of finite temperature QCD plasma has attracted
a lot of study in the past few years. One of the interesting features is that the plasma produced at RHIC
seems to be strongly interacting, and perturbative QCD study can't explain certain phenomena observed,
such as jet-quenching and the small viscosity-to-entropy ratio, etc.
Given the situation, gauge/gravity correspondence may be of some help to understand at least
certain aspects which are common to a wide class of examples of strongly interacting gauge theories
at finite temperature \cite{Nastase:2005rp}.
A slightly simpler version of conformal field theories, the AdS/CFT correspondence,
always contains a 5D Einstein gravity with cosmological constant in asymptotic $AdS_5$ spacetime,
and the predictions from it will be universal ones. The hope is that real QCD plasma shares some of these predictions.
Besides that, it is also of purely theoretical interest to study finite temperature plasma dynamics in the dual gravity picture.
In the gravity dual picture, a finite temperature plasma {\it in equilibrium} is described by
a black-brane, a still mysterious object, whose Hawking temperature is identified as the temperature of the field theory plasma \cite{Witten:1998zw},
and the small near-equilibrium gravity dynamics on this black-brane spacetime reproduces
the expected hydrodynamic description of a finite temperature gauge theory plasma \cite{Policastro:2002se}.
The rewards are various transport coefficients of strongly interacting gauge theories near equilibrium, which
cannot be computed by current field theory techniques \cite{Policastro:2001yc,Buchel:2003tz}.

Although the RHIC plasma is believed to be quickly local-thermalized to facilitate a hydrodynamic description
\cite{Teaney:2001av}, it is {\it not} a near-equilibrium plasma; it is an expanding plasma with continual
decrease of temperature, whose ultimate state would be a cold gas of hadrons. Only the late stage of
asymptotically slow variation of temperature can be described by hydrodynamics. Therefore, the black-brane for a
plasma at equilibrium is not {\it a priori} correct object one can start with, and one needs to look for a full
non-linear gravity solution to describe an expanding plasma. The early time dynamics of the collision, such as
isotropization \cite{Chesler:2008hg}, would require a hard-core numeric analysis of Einstein equation
\cite{Kovchegov:2007pq,Grumiller:2008va,Gubser:2008pc,Bhattacharyya:2009uu}, but if one is interested in only
late time asymptotic regime, a systematic perturbative approach in terms of suitably chosen expansion parameters
may be invoked to find a solution. This was first initiated by Janik-Peschanski \cite{Janik:2005zt} with several
notable developments afterward in
Ref.\cite{Nakamura:2006ih,Benincasa:2007tp,Alsup:2007bs,Buchel:2008ac,Heller:2008mb,Kinoshita:2008dq,Buchel:2008kd}.

A useful assumption in studying the relevant expanding plasma is the boost-invariance first proposed by Bjorken \cite{Bjorken:1982qr},
which simplifies the gravity analysis drastically. It has a well-accepted physics motivation from what would happen in
the collision of two heavy nuclei. We refer the readers to many nice reviews on that, for example Ref.\cite{Heller:2008fg}.
To explain kinematics in few words, take the coordinate transformation to
\be
x^0=\tau {\rm cosh} y\quad,\quad x^3=\tau {\rm sinh} y\quad,
\ee
where $(x^0,x^3)$ are the Minkowski time and the collision direction respectively.
The flat metric then looks as
\be
ds^2=-d\tau^2+\tau^2 dy^2+\sum_{i=1}^2 dx_i^2\quad.\label{rindler}
\ee
A boost transformation along $x^3$ is simply a constant shift of rapidity $y$, and the boost-invariance
implies the homogeneity along $y$. For simplicity one also assumes independence of transverse directions $x^{1,2}$,
which is ok at least in the region nearby central collision axis,
so that all physical quantities would depend on the proper-time $\tau$ only. For the hydrodynamic plasma,
this means that the energy-momentum tensor is written as $T^{\mu\nu}(\tau)$.
One more thing, which is again well-motivated in collisions of two identical projectiles, is the symmetry
under the flipping of spacetime rapidity $y$,
\be
P\quad:\quad y \leftrightarrow -y\quad,
\ee
which is also a space parity transformation $x^3\leftrightarrow -x^3$.
This assumption would put the $\tau y$-component of the energy-momentum tensor zero; $T^{\tau y}(\tau)\equiv 0$,
or equivalently the local fluid component has a 4-velocity $u^\mu$ with $u^\tau=1$ being the only non-zero component.
Given these inputs, the corresponding gravity dual solution for the boost invariant plasma has been analyzed extensively.

Our aim is to add a slight twist to the above by allowing components that are odd under the parity flipping of rapidity $y$,
while still taking boost-invariance for granted; simply put, our plasma of interest will be $y$-homogeneous with non-zero $T^{\tau y}(\tau)\neq 0$.
This would complicate the local fluid velocity $u^\mu$ with a non-zero $u^y\neq 0$, whose value is not clear {\it a priori}
and should be determined along the way in the subsequent analysis.
We stress that the existence of $T^{\tau y}(\tau)\neq 0$ has nothing to do with the total momentum along $x^3$ ($P^3$) in the original
Minkowski coordinate which can always be made zero by suitable boost transformations going to the center of mass frame.
An easiest way of seeing this is to recall that boost-transformation is a simple shift in $y$, under which $T^{\tau y}(\tau)\neq 0$ is invariant.
The $P^3$ is in fact given by
\be
P^3 = \int d\left(\tau \sinh y\right)\,\tau(1+2\sinh^2 y) T^{\tau y}\quad,
\ee
so that the shift in $y$ indeed affects it, but the statement that there is a region with a boost-invariant $T^{\tau y}(\tau)\neq 0$
is something which is not related to whether one is at the center of mass frame or not.
Rather, the region with $T^{\tau y}(\tau)\neq 0$ has an intrinsic asymmetry that presumably
arises from two different species of projectiles, such as lead-gold collision for example. See Figure 1 for a schematic explanation.
\begin{figure}[t]
\begin{center}
\includegraphics[width=12cm]{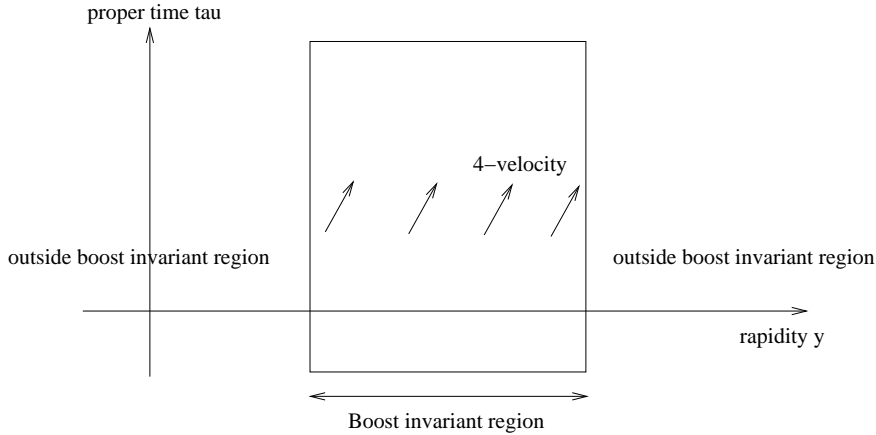}
\caption{A schematic picture for parity asymmetric boost invariant plasma.}
\label{d1}
\end{center}
\end{figure}
It would be interesting to think further whether this is a realistic possibility in experiments\footnote{ Near
the completion of this work, we became aware of Ref.\cite{Alsup:2006nc} which considered R-charged plasma with
leading order parity asymmetric term, Ref.\cite{Kajantie:2007bn} which considered a similar construction in 1+1
dimensions, and Ref.\cite{Albacete:2009ji} which studied asymmetric collision of two shock waves in $AdS_5$.}.

Besides the question of reality of our parity-asymmetric boost invariant plasma,
there is another motivation to study it, perhaps more important in the context of AdS/CFT correspondence.
The gravity solution we will construct is one new time-dependent gravity background,
whose late time asymptotic behaviors can provide us consistent cross-checks for
the recently proposed second order conformal viscous hydrodynamics in Ref.\cite{Baier:2007ix,Bhattacharyya:2008jc}.
Especially, the second order transport coefficients have been obtained by several other methods,
and it would be a non-trivial corroboration of the framework if one finds agreement between them and our results,
which is indeed the case as we will see.
Finally, we should mention that our gravity analysis is guided by previous works in a large extent, especially those in Ref.\cite{Heller:2008mb,Kinoshita:2008dq}.
In most cases, we will follow the conventions and notations in Ref.\cite{Kinoshita:2008dq}.

\section{Precursory hydrodynamics}

We will be back to the full second order conformal hydrodynamics in section 5, but
before presenting the gravity analysis in the next section, it is helpful to
consider a preliminary hydrodynamics to see what one would expect from the additional parity-asymmetric term we are putting.
The important thing we will find is that its first non-trivial effects appear in the usual parity-symmetric terms
at second order in late time expansion.

The conservation equations $D_\mu T^{\mu\nu}=0$ in our Rindler-like coordinate (\ref{rindler}) are
\bear
\partial_\tau \left(\tau T^{\tau\tau} \right)+\tau^2 T^{yy} & =&  0\quad,\nonumber\\
\partial_\tau \left(\tau T^{\tau y} \right) + 2 T^{\tau y} &=& 0 \quad,\nonumber\\
\partial_\tau \left( \tau T^{\tau i}\right) &=& 0\quad,
\eear
and the traceless condition is
\be
-T^{\tau\tau} +\tau^2 T^{yy} + T^{11}+T^{22} =0\quad.
\ee
One easily integrates the second equation to get the parity-odd term
\be
T^{\tau y} = -{4\over 3}{C\over \tau^3}\quad,
\ee
with an integration constant $C$ which we will keep non-zero. We do abandon $T^{\tau i}$ however in this work.
The remaining equations are as usual as in the parity-symmetric
cases, which one can solve in terms of a single function $a^{(4)}(\tau)$;
\bear
T^{\tau\tau}&=& -a^{(4)}(\tau)\quad,\nonumber\\
T^{yy}&=&{1\over \tau^2} \left(a^{(4)}(\tau)+\tau \partial_\tau a^{(4)}(\tau)\right)\quad,\nonumber\\
T^{ii}&=& -{1\over 2}\left(2 a^{(4)}(\tau) +\tau\partial_\tau a^{(4)}(\tau)\right)\quad,i=1,2\quad,\label{general}
\eear
where the notation is chosen for later convenience. The function $a^{(4)}(\tau)$ contains all the
non-trivial dynamical details of the expanding plasma, and depends on the microscopic theory as well.
Note that the parity-asymmetric component $T^{\tau y}\neq 0$ has a fixed, trivial structure $\sim {1\over \tau^3}$ due to kinematics,
but its effects on $a^{(4)}(\tau)$ may well be highly non-trivial; in fact there is no {\it a priori}
expectation one can make except that it should be an even function on $C$. There is no further information one can extract from
the above kinematic constraints.

One might be misled to think that $a^{(4)}(\tau)$
would be simply decoupled from $T^{\tau y}$, but it is not
true in general as the plasma is interacting non-linearly.
To see this clearly, let's try to map the energy-momentum tensor into an ideal hydrodynamics form, which is only a crude
leading order approximation in the late time expansion but would be sufficient to make the point here,
\be
T^{\mu\nu}= \epsilon(\tau) u^\mu u^\nu + p(\tau) \Delta^{\mu\nu} +{\rm ( higher\,\,derivative\,\,terms )}\quad, \Delta^{\mu\nu}=g^{\mu\nu}+u^\mu u^\nu\quad,\label{ideal}
\ee
where $p(\tau)={1\over 3} \epsilon(\tau)$ for conformal plasmas. For parity-symmetric boost invariant plasma with $u^\tau=1$
being the only component in $u^\mu$, one can insert the above into (\ref{general}) to get the equation
\be
\epsilon = -{3\over 4} \tau \partial_\tau \epsilon +\cdots \quad,
\ee
with a solution as the leading order behavior in large $\tau$, (we have put $\epsilon_0\equiv 1$ for simplicity)
\be
\epsilon(\tau) \sim {1\over \tau^{4\over 3}} +\cdots\quad.
\ee
Then consider our situation of having a non-zero $T^{\tau y}$,
and an inspection requires us to have a non-zero $u^y$ from
\be
T^{\tau y}={4\over 3}\epsilon(\tau)u^\tau u^y + \cdots\quad.\label{uy}
\ee
Assuming that the leading large $\tau$ behaviors of $\epsilon(\tau)$ and $u^\tau$ are unaffected by the parity-asymmetric term,
\be
\epsilon(\tau)\sim {1\over \tau^{4\over 3}} +\cdots\quad,\quad u^\tau=1+\cdots\quad,
\ee
which indeed will turn out to be true consistently in later analysis,
one obtains from (\ref{uy}) that
\be
u^y\sim -{C\over \tau^{5\over 3}}+ \cdots\quad,
\ee
as the first non-vanishing term in the late time expansion.
From the normalization $-1=g_{\mu\nu}u^\mu u^\nu =-(u^\tau)^2 + \tau^2 (u^y)^2$, one then finds the sub-leading
correction to $u^\tau$ to be,
\be
u^\tau=1+ {C^2\over 2 \tau^{4\over 3}}+ \cdots\quad.
\ee
Back to (\ref{ideal}) with this $u^\mu$, one finds that
sub-leading corrections coming from the parity-asymmetric term to the parity-symmetric components of the energy momentum tensors are
$\sim {1\over \tau^{4\over 3}}$ smaller than the leading order terms.
This is a second order correction in late time expansion with ${1\over \tau^{2\over 3}}$.
Although we need to consider higher order viscous terms that we neglected in (\ref{ideal}) to do a proper analysis at this second order,
which will be the subject in section 5, this preliminary analysis shows that
non-trivial effects from the parity-asymmetric component $T^{\tau y}$ do appear in $a^{(4)}(\tau)$
starting at  second order in the late time expansion of ${1\over \tau^{2\over 3}}$.
This input will be helpful to the gravity analysis in the next section.

\section{Gravity solution}

Any AdS/CFT set-up allows a consistent truncation of the 5D bulk dual theory to a pure Einstein gravity with
negative cosmological constant. As there is no global charge in the plasma we are considering, we work within
this gravity sector only. See Ref.\cite{Bak:2006dn,Erdmenger:2008rm,Torabian:2009qk,Cardoso:2009mt} for
R-charged plasmas. We take the standard convention of putting the $AdS_5$ radius $\ell\equiv 1$ and the
cosmological constant $\Lambda=-6$. This effectively fixes the 5D Planck constant to be a specific value
depending on the model. For ${\cal N}=4$ SYM, it is \be {8\pi G_5}= {4\pi^2\over N_c^2}\quad. \ee The equation
of motion we have to solve for the dual geometry of expanding plasma is\footnote{Capital letters are for 5D
coordinates, while Greek letters are used for 4D indices.} \be E_{MN} \equiv R_{MN}- {1\over 2} R \,g_{MN} -6
g_{MN}= R_{MN} +4 g_{MN} =0\quad.\label{einstein} \ee Our subsequent analysis is largely based on the previous
works starting from Ref.\cite{Janik:2005zt} where the late time expansion scheme of $1\over \tau^{2\over 3}$ was
first proposed in the Fefferman-Graham Ansatz. It was later found in Ref.\cite{Heller:2008mb,Kinoshita:2008dq}
that an Eddington-Finkelstein form of the metric is more suitable for all order regularity of the solution, and
this will be our starting Ansatz too. Our notations and methods closely follow those in
Ref.\cite{Kinoshita:2008dq}.

We propose the following metric solving the 5D einstein equation,
\bear
ds^2 &=& -r^2 a(r,\tau) d\tau^2 + 2 d\tau dr +e^{2\left(b(r,\tau)-c(r,\tau)\right )}\left(1+r\tau\right)^2 dy^2
+r^2 e^{c(r,\tau)}\sum_{i=1}^2 dx_i^2
\nonumber\\
&+& 2r\left(F(r,\tau) + {h(r,\tau)\over 3 r \tau}\right)\left(1+r \tau\right) d\tau dy + {2\over r} \left(1+r\tau\right) h(r,\tau) dy dr\quad,
\eear
where the second line is what we add as new terms violating parity symmetry $y\leftrightarrow -y$.
The radial coordinate $r$ is the 5'th holographic direction in addition to the previous 4D Rindler-like coordinates $(\tau,y,x_1,x_2)$,
and $r\to \infty$ is the $AdS_5$ boundary. The factors of $\left(1+ r\tau\right)$ are just for convenience and not essential
in late time expansion, but we choose to put them according to Ref.\cite{Kinoshita:2008dq}.
Near $r\to\infty$ boundary, the induced 4D metric should converge to $r^2 \left(-d\tau^2+\tau^2 dy^2+dx_i^2\right)$ in order to be asymptotic
$AdS_5$,
so that the functions that appear in the above have boundary conditions
\be
(a,b,c,F)\to (1,0,0,0)\quad,\quad r\to\infty\quad.\label{bdy}
\ee

A crucial element in the consistent late time expansion for the gravity solution
is to introduce a scaling variable
\be
u\equiv r\tau^{1\over 3}\quad,
\ee
that is kept finite while taking a late time limit $\tau\to\infty$, and to invoke a power series
expansion of $1\over\tau^{2\over 3}$ for every functions to be determined in the metric by solving
the einstein equation order by order \cite{Janik:2005zt}. To be more concrete,
\bear
a(r,\tau) &=& a_0(u)+{a_1(u)\over\tau^{2\over 3}}+{a_2(u)\over \tau^{4\over 3}} + \cdots\quad,\nonumber\\
b(r,\tau) &=& b_0(u)+{b_1(u)\over\tau^{2\over 3}}+{b_2(u)\over \tau^{4\over 3}} + \cdots\quad,\nonumber\\
c(r,\tau) &=& c_0(u)+{c_1(u)\over\tau^{2\over 3}}+{c_2(u)\over \tau^{4\over 3}} + \cdots\quad,\nonumber\\
F(r,\tau) &=& F_0(u)+{F_1(u)\over\tau^{2\over 3}} + \cdots\quad,\nonumber\\
h(r,\tau) &=& h_0(u)+{h_1(u)\over\tau^{2\over 3}} + \cdots\quad,\label{exp}
\eear
where we have shown terms only up to the point of our interest for the purposes of this work,
but in principle one can go further as he/she wants.
It is conceptually straightforward to put the expansion (\ref{exp}) into the einstein equation (\ref{einstein}) and to solve
order by order iteratively. We will find that this works consistently
with our metric Ansatz.

Although it is conceptually simple, it is technically demanding to compute the relevant einstein tensors with our metric,
and we have used a freeware Mathematica package, RIEMANNIAN GEOMETRY $\&$ TENSOR CALCULUS \cite{mathematica},
for algebraic computations. However, what will be important for us at the end is a closed form of the solution up to the desired order
that we will present in this section.
One subtle point which is worth of mentioning is that
solving $E_{MN}=0$ at a given order is {\it not} equivalent to solving $E^M_N=0$ at the same order,
because various metric coefficients contain different factors of $\tau$, so that in
going from $E_{MN}=0$ to $E^M_N=0$, the terms at different orders in the former can mix up at a same order in the latter.
Because the latter seems to be a more invariant notion under the late time scaling $\tau\to\infty$
(recall that $g^M_N=\delta^M_N\sim{\cal O}(1)$ for a diagonal metric), we will work with it.

Let us denote the coefficient of $\tau^{-\alpha}$ in the einstein equation $E^M_N=0$ as $E^{M(\alpha)}_N$.
For the zero'th order functions $(a_0(u),b_0(u),c_0(u),F_0(u),h_0(u))$, one
obtains several {\it non-linear} equations of them from the leading order terms of the einstein equation.
There is no definite way of solving these; it is essentially same to finding a solution of non-linear einstein gravity.
For our purpose, one can proceed by the observation in the previous section that the parity-asymmetric terms start
to affect parity-symmetric components at second order in late time expansion. This suggests that
one can simply take the previously known solution for $(a_0(u),b_0(u),c_0(u))$ without parity-asymmetry terms,
while putting $(F_0(u),h_0(u))=(0,0)$ as a leading trial solution;
\be
a_0(u)=1-{w^4\over u^4}\quad,\quad b_0(u)=c_0(u)=F_0(u)=h_0(u)=0\quad.
\ee
One easily checks that this solves the leading order einstein equations
up to $E^{M(0)}_N$, and it can be a consistent starting point to do subsequent series expansion.
Once the zero'th order solution is given, the problem becomes much more easier and systematic, because
the next leading einstein equations give {\it linear} differential equations for the next order
terms in the solution $(a_1(u),b_1(u),c_1(u),F_1(u),h_1(u))$, and so on.
One can in principle go on further order by order  iteratively.

One finds that the next order einstein equations, $E^{\tau({2\over 3})}_\tau$, $E^{\tau({1\over 3})}_u$, $E^{u({5\over 3})}_\tau$,
$E^{u({2\over 3})}_u$, $E^{y({2\over 3})}_y$, and $E^{i({2\over 3})}_i$
provide complete linear differential equations for $(a_1(u),b_1(u),c_1(u))$ only, without
involving $(F_1(u),h_1(u))$ at all. 
Therefore, their solution is simply identical to the previous results of the parity-symmetric case,
\bear
a_1(u)&=&{2\over 3}{w^3(u+w)+C_1(u^4+w^4)\over u^5}\quad,\nonumber\\
b_1(u)&=& {C_1\over u}\quad,\\
c_1(u)&=& {1\over 6w }\left(-\pi  +4(1+C_1){w\over u}+2\tan^{-1}\left({u\over w}\right)
+ \log\left(u^4\over (u+w)^2 (u^2+w^2)\right)\right)\quad,\nonumber
\eear
where $C_1$ is an integration constant\footnote{In comparing with Ref.\cite{Kinoshita:2008dq}, $C_1=-\xi_1-1$.} which is actually a pure coordinate reparametrization freedom,
and we have already determined another integration constant to have a regularity at $u=w$
which is the only dangerous point except the trivial singularity at $u=0$ hidden behind event horizon.
This is indeed in line with our expectation from the previous section that non-trivial effects
of parity-asymmetry would emerge starting only at the second order.

One obtains differential equations for $(F_1(u),h_1(u))$ from the parity-odd components of the einstein equation. From $E^{\tau(-{1\over 3})}_y$, $E^{y({4\over 3})}_u$, and $E^{y({5\over 3})}_\tau$ we get
\be
u F_1''(u) +5 F_1'(u)=0\quad,
\ee
whose solution with the boundary condition (\ref{bdy}) is
\be
F_1(u)={C\over u^4}\quad.
\ee
The remaining component, $E^{u({4\over 3})}_y$, gives us an equation for $h_1(u)$, but
{\it only} in the combination of
\be
S_1(u)\equiv h_1(u)+{3\over 4 }u^2 F_2'(u)\quad.
\ee
In fact, if we look at the next order parity-odd einstein equations,
$E^{\tau({1\over 3})}_y$, $E^{y({2})}_u$, $E^{y({7\over 3})}_\tau$, they are equations for $S_1(u)$
without separate $h_1(u)$ or $F_2(u)$. All these equations are solved uniquely by
\be
S_1(u)=C {u \left(21 u + 10 C_1\right)\left( u^2 +uw+w^2\right) + 5\left(5+2C_1\right )w^3\over 4 u^4 \left(u+w\right)\left(u^2 +  w^2\right)}\quad.
\ee
The reason for this combination can be traced back to the fact that the $F_2(u)$ can be
simply gauged away by the coordinate transform
\be
y\to y+{3\over 4} F_2(u){1\over\tau^{4\over 3}}+\cdots \quad,
\ee
under which $h_1(u)$ becomes $h_1(u)+{3\over 4 }u^2 F_2'(u)\equiv S_1(u)$, so that only $S_1(u)$
is a gauge invariant combination at this order. This means that one can choose the gauge
$F_2(u)\equiv 0$, which gives finally
\be
h_1(u)=C {u \left(21 u + 10 C_1\right)\left( u^2 +uw+w^2\right) + 5\left(5+2C_1\right )w^3\over 4 u^4 \left(u+w\right)\left(u^2 +  w^2\right)}\quad.
\ee
It seems to us that we can always gauge away $F_n(u)$, $n\ge 2$ with similar coordinate
reparametrizations of $y$ order by order, so that there is a gauge with $F(u,\tau)$ being
\be
F(u,\tau)={C\over u^4}{1\over \tau^{2\over 3}}\quad,
\ee
{\it exactly} without any higher-order modifications. This seems to be consistent with the fact
in the previous section that $T^{\tau y}$ has a fixed structure of $1\over\tau^{3}$
without modifications due to kinematics.

Having obtained $(F_1(u),h_1(u))$, we then consider the next order einstein equations
for $(a_2(u),b_2(u),c_2(u))$, and one indeed discovers non-trivial effects from
$(F_1(u),h_1(u))$ at this order.
$E^{\tau({4\over 3})}_\tau$, $E^{\tau({1})}_u$, $E^{u({7\over 3})}_\tau$,
$E^{u({4\over 3})}_u$, $E^{y({4\over 3})}_y$, and $E^{i({4\over 3})}_i$ provide
complete differential equations for $(a_2(u),b_2(u),c_2(u))$
which include "sources" from various quadratic combinations of $(F_1(u),h_1(u))$,
in addition to the usual terms from parity-symmetric components. One can also check that
there are no further contributions to these equations from higher order terms in the expansion.
We simply present the solution of these equations
after fixing certain integration constants to be regular at $u=w$,
\bear
a_2(u)&=&{1\over 3u^5}\Bigg({1\over w}\left(u^4+w^4\right)\tan^{-1}\left({u\over w}\right)
-uw^2\left({1\over 2}+{\log 4\over 6}+\log\left({u^2\over u^2+w^2}\right)\right) \nonumber\\
&+& 1+{C_1\over 3}\left(6+C_1\right) u^3 -2 C_2 u^4 -(1+C_1)^2 {w^4\over u}-2 w^3\left(
1+{2 C_1\over 3}+C_2 w\right)\Bigg)\nonumber\\
&-& {C^2\over u^{16} w^6}\Bigg(
{u^{11}\over w}\left(u^4+w^4\right)\tan^{-1}\left({u\over w}\right)+u^{14}+{1\over 3}u^{12} w^2 +{6\over 5} u^{10} w^4+{14\over 45} u^6 w^8 +{1\over 9} u^2 w^{12}\nonumber\\
&-& {25\over 308}\left(5+2C_1\right)^2 w^6
\left({11\over 4} u^8+u^4 w^4 -{4\over 7} w^8\right)
-{\left(5+2C_1\right)u w^9\over 4}\left(u^4+w^4\right)\Bigg)\quad,\\
b_2(u)&=&{\pi\over 12 w^2}+{1\over 2w}\left({1\over u}-{1\over 3w}\right)\tan^{-1}\left({u\over w}\right)
+{1\over 6w^2}\left(\log\left({u+w\over u^4}\right) +{3\over 2}\log\left(u^2+w^2\right)\right)\nonumber\\
&-&{1\over 3u w}\left(1+{w\over u}\left(-1+C_1\left(1+{C_1\over 2}\right)+3 C_2 u\right)\right)\nonumber\\
&-&C^2\bigg({3\over 2 u w^7}\tan^{-1}\left({u\over w}\right)
+{1\over 2 u^2}\left({3\over w^6}-{1\over u^2 w^4}+{3\over 5 u^4 w^2}+{w^2\over 3 u^8}\right)\nonumber\\
&-&{75\over 224}{\left(5+2C_1\right)^2\over u^8}\left(1-{7w^4\over 11 u^4}\right)+{3\left(5+2C_1\right)\over
8}{w^3\over u^{11}}\Bigg)\quad,\\
c_2'(u)&=&
-{1\over 9u^2 w (u^4-w^4)}\Bigg(
\left(u^4+2u w^3-3 w^4\right)\tan^{-1}\left({u\over w}\right)+u\bigg(u^3\log\left({(u+w)^2 (u^2+w^2)\over u^4}\right)\nonumber\\&+& w^3\log\left({(u^2+w^2)\over 4(u+w)^2}\right)\bigg)\Bigg)
-{1\over 9 u^3 w (u-w) (u+w)^2(u^2+w^2)^2}\Bigg(\pi u^8 -u^7 w-u^6 w^2 \nonumber\\ &-&(3+\pi)u^5 w^3
-(7+2\pi +10C_1)u^4 w^4 +2(2+C_1) u^3 w^5 +2(2+C_1)u^2 w^6 \nonumber\\
&+&(2+2C_1+\pi)u w^7 + (2+4C_1+\pi) w^8+(\pi-2C_1^2 -6 C_2 u) w (u-w)\bigg( u^6+2 u^5 w\nonumber\\
&+& 3u^4 w^2 +4 u^3 w^3 +3 u^2 w^4 +2 u w^5 +w^6 \bigg)\Bigg)\nonumber\\
&+&C^2\Bigg({1\over u^2 w^7}\tan^{-1}\left({u\over w}\right)
+{1\over u^{13} w^6 (u^2+w^2)}\bigg(u^{12}+2 u^{10} w^2 +{6\over 5} u^8 w^4
-{3041\over 70}u^6 w^6\nonumber \\ &-&{5485\over 126} u^4 w^8 +{17315\over 396} u^2 w^{10}
-{50\over 7} C_1 (5+C_1)w^6\left(u^6+u^4 w^2-{21\over 22} u^2 w^4\right)\nonumber\\
&-&{11\over 4}(5+2C_1) u w^9 (u^2+w^2)+{75\over 44}(5+2C_1)^2 w^{12} \bigg)\Bigg)\quad,
\eear
where $C_2$ is a constant of integration\footnote{It is similar to $\xi_2$ in Ref.\cite{Kinoshita:2008dq}.},
which is a pure gauge. Note that the pieces proportional to $C^2$ are the new terms
coming from our parity-asymmetric components.

Up to this order, we find the unique regular solution except $C_1$ and $C_2$ which
correspond to a coordinate transformation as observed in Ref.\cite{Kinoshita:2008dq},
\be
  u\to u+{C_1\over 3 \tau^{2\over 3}} -{C_2\over 3\tau^{4\over 3}} +\cdots\quad,
\ee
and there is no further ambiguity left in the solution.
It seems very plausible that all order regularity in the present expansion scheme
can also be proved.

\section{Holographic renormalization and energy momentum tensor}

Our next obvious step is to find the 4D energy-momentum tensor of the
parity-asymmetric expanding plasma corresponding to the
gravity solution we constructed in the previous section via AdS/CFT correspondence.
This involves
the well-established procedure of holographic renormalization \cite{Balasubramanian:1999re,de Haro:2000xn}, and we will be very brief
in explaining it.
Consider a constant, large $r$ hyper-surface as a boundary of our asymptotic $AdS_5$ spacetime,
whose induced metric we denote as $\gamma_{\mu\nu}$.
Let $K_{\mu\nu}$ be the extrinsic curvature of the hyper-surface.
The "bare" energy-momentum tensor computed from the well-known Brown-York method
for the spacetime with boundary,
\be
T_{\mu\nu}^{\rm bare} = {1\over 8\pi G_5} r^2 \left( K_{\mu\nu} - K \gamma_{\mu\nu}\right)\quad,
\ee
turns out to be divergent in $r\to\infty$ limit.
The factor $r^2$ in front is from the fact that $\gamma_{\mu\nu}=r^2 g_{\mu\nu}^{\rm CFT}$, so
\be
{1\over\sqrt{-g^{\rm CFT}}}{\delta\over \delta g^{\mu\nu}_{\rm CFT}}
= r^2 {1\over\sqrt{-\gamma}}{\delta\over \delta \gamma^{\mu\nu}}\quad.
\ee
This divergence has a field theory interpretation
of UV divergences one encounters for the "bare" operators before regularization and renormalization.
Therefore the question is what are the correct counter-terms one should subtract
from the above to have a well-defined renormalized expectation value of $T_{\mu\nu}$.
The important ingredient in choosing these counter-terms is
the requirement of general covariance to correctly reproduce certain conformal anomalies.
The upshot is that the counter-terms should be local covariant objects constructed purely out of the
induced metric $\gamma_{\mu\nu}$ on the boundary. In a "minimal subtraction scheme", one choose them minimally just to be sufficient to remove all the divergences in the bare terms.
The result is
\be
T_{\mu\nu}={1\over 8\pi G_5} {\rm lim}_{r\to\infty} r^2  \left(K_{\mu\nu} - K \gamma_{\mu\nu}
-3\gamma_{\mu\nu} +{1\over 2} G_{\mu\nu}\right)\quad,\label{EMtensor}
\ee
where $G_{\mu\nu}=R_{\mu\nu}-{1\over 2}R \,\gamma_{\mu\nu}$ is the einstein tensor of $\gamma_{\mu\nu}$.

It might look rather straightforward to implement this procedure by simply plugging the gravity
solution of the previous section into (\ref{EMtensor}), but in reality one finds it
to be impossibly complicated to do in this way.
A better strategy that has been used in practice all the time, and also in the formulation
of holographic renormalization itself, is
to first solve the 5D einstein equation near boundary $r\to\infty$ {\it in general},
reducing the number of independent degrees of freedom, and to use these results
in computing the energy-momentum tensor (\ref{EMtensor}).
One then simply needs to
identify the relevant near-boundary terms, that appear in the energy-momentum tensor expression,
from the actual gravity solution to get final results.
In obtaining near-boundary form of the metric that solves the einstein equation,
we will use one fact from the previous section; note that $F_1$ and $h_1$
in our solution have ${1\over u^4}\sim {1\over r^4}$ behavior near $r\to\infty$, so that up to our desired order $F$ and $h$ can be safely assumed to have expansion starting
\be
F(r,\tau) = {F^{(4)}(\tau)\over r^4}+\cdots\quad,\quad
h(r,\tau) = {h^{(4)}(\tau)\over r^4}+\cdots\quad.\label{fh}
\ee
We also checked independently that these are  consistent with the einstein equation near boundary.
Because the parity-symmetric components $a(r,\tau),b(r,\tau),c(r,\tau)$ are
even functions on $(F,h)$, the effects from $(F,h)$ to $(a,b,c)$ can be at most $1\over r^8$.
As the energy-momentum tensor is sensitive only up to $1\over r^4$ terms, these effects
won't affect the near-boundary {\it structure} of $(a,b,c)$ that are
relevant for the energy-momentum tensor. We stress that this {\it does not}
mean that the {\it values} of $(a,b,c)$  up to $1\over r^4$ near boundary, and hence the energy-momentum tensor result, is not affected
by the presence of $(F,h)$. Only the near-boundary structure is insensitive, whose meaning
will be clearer in a moment.
By assuming (\ref{fh}), explicit computations give us
\bear
a(r,\tau)&=& 1+{a^{(1)}(\tau)\over r}+ \left(
{\left(a^{(1)}(\tau)\right)^2\over 4}-\partial_\tau a^{(1)}(\tau)\right){1\over r^2}+
{a^{(4)}(\tau)\over r^4}+\cdots\quad,
\nonumber\\
e^{2(b(r,\tau)-c(r,\tau))}&=& 1
+{a^{(1)}(\tau)\over r}+\left(
{\left(a^{(1)}(\tau)\right)^2\over 4}-{ a^{(1)}(\tau)\over\tau}\right){1\over r^2}
+\left(
-{\left(a^{(1)}(\tau)\right)^2\over 2\tau}+{ a^{(1)}(\tau)\over\tau^2}\right){1\over r^3}\nonumber\\
&+&\left(
{3\left(a^{(1)}(\tau)\right)^2\over 4\tau^2}-{ a^{(1)}(\tau)\over\tau^3}
+a^{(4)}(\tau)+{3\over 4}\tau\partial_\tau a^{(4)}(\tau)\right){1\over r^4}+\cdots\quad,\\
e^{c(r,\tau)}&=& 1+ {a^{(1)}(\tau)\over r}
+
{\left(a^{(1)}(\tau)\right)^2\over 4 r^2}+
\left(
-{1\over 2} a^{(4)}(\tau) -{3\over 8}\tau\partial_\tau a^{(4)}(\tau)\right){1\over r^4}+\cdots\quad,\nonumber
\eear
which is essentially identical to those in Ref.\cite{Kinoshita:2008dq} without $(F,h)$ as mentioned before.
However, the function $a^{(4)}(\tau)$ which is undetermined by boundary analysis {\it does} depend on $(F,h)$
or equivalently $C$. Finally, $h^{(4)}(\tau)$ can be given in terms of $F^{(4)}(\tau)$ by einstein equation,
but we won't need it for our purposes.

With the above, the renormalized energy-momentum tensor is easily found to be
\bear
T_{\tau\tau}&=&{1\over 8\pi G_5}\left(-{3\over 2} a^{(4)}(\tau)\right)\quad,\nonumber\\
T_{\tau y}&=&{1\over 8\pi G_5}\left(2\tau  F^{(4)}(\tau)\right)\quad,\nonumber\\
T_{yy}&=&{1\over 8\pi G_5}\left({3\over 2}\tau^2\left( a^{(4)}(\tau)+\tau\partial_\tau a^{(4)}(\tau)\right)\right)\quad,\nonumber\\
T_{ii}&=&{1\over 8\pi G_5}\left(-{3\over 4}\left( 2a^{(4)}(\tau)+\tau\partial_\tau a^{(4)}(\tau)\right)\right)\quad,i=1,2\quad.\label{grav1}
\eear
Note that $a^{(1)}(\tau)$ doesn't appear because it can be removed by a coordinate transformation.
From the explicit solution in the previous section, one can easily obtain the required
$a^{(4)}(\tau)$ and $F^{(4)}(\tau)$ by recovering $r=u \tau^{-{1\over 3}}$,
\bear
a^{(4)}(\tau)&=&-{w^4\over \tau^{4\over 3}}
+{2 w^3\over 3 \tau^2} -{(1+2 \log 2)w^6 +12 C^2 \over 18 w^4 \tau^{8\over 3}} +\cdots\quad,\nonumber\\
F^{(4)}(\tau) &=& {C\over \tau^2}\quad,\label{result}
\eear
where we can see the parity-asymmetric effects of $C$ at the second order expansion.

\section{Mapping to second order conformal viscous hydrodynamics}

In this section, we will try to map the energy-momentum tensor from the gravity side in the previous section to the recently proposed second-order conformal viscous hydrodynamics in Ref.\cite{Baier:2007ix,Bhattacharyya:2008jc}.
We remind the readers of that this is not a priori guaranteed to work because the proposed hydrodynamics
framework is restrictive in describing the energy-momentum of plasma, with
certain finite number of transport coefficients. Any success in this mapping will be a corroboration
of the validity of the framework, and we think this is at least worth of checking explicitly.
We will find that
the mapping indeed works with some second order transport coefficients identified along the way,
whose results are in perfect agreement with those in the literature via other methods/backgrounds.

Simply put, the second order conformal viscous hydrodynamics (SCVH) in Ref.\cite{Baier:2007ix} is
a statement about the energy-momentum tensor of plasma, sometimes called {\it constitutive relations},
expressed in terms of a finite set of local thermodynamics quantities. In our case at hand,
these will be 4-velocity $u^\mu$ with $u_\mu u^\mu =-1$ and the energy density $\epsilon$.
The pressure $p={1\over 3}\epsilon$ is dictated by $\epsilon$ by conformal nature.
There are four unknowns and we have four equations from $D_\mu T^{\mu\nu}=0$, so that
one is given a complete, closed system of dynamics.
The proposal was
\bear
T^{\mu\nu}&=& \epsilon u^\mu u^\nu +p \Delta^{\mu\nu}
-\eta \sigma^{\mu\nu} +\eta\tau_{II}\left[ ^\langle D\sigma^{\mu\nu\rangle}+{1\over 3} \sigma^{\mu\nu}\left(\nabla\cdot u\right)\right]\nonumber\\
&+&\kappa\left[R^{\langle \mu\nu \rangle}-2 u_\alpha R^{\alpha\langle \mu\nu \rangle \beta}
u_\beta\right]+\lambda_1 \sigma^{\langle \mu}_{\,\,\,\,\,\lambda} \sigma^{\nu\rangle \lambda}
+\lambda_2 \sigma^{\langle \mu}_{\,\,\,\,\,\lambda} \Omega^{\nu\rangle \lambda}
+\lambda_3 \Omega^{\langle \mu}_{\,\,\,\,\,\lambda} \Omega^{\nu\rangle \lambda}\quad,\label{proposal}
\eear
where $\Delta^{\mu\nu}=g^{\mu\nu}+u^\mu u^\nu$ is the projection to the transverse space to $u^\mu$,
$D=u^\mu \nabla_\mu$, and
\be
\sigma^{\mu\nu}=2 \nabla^{\langle \mu} u^{\nu \rangle}\quad,\quad
\Omega^{\mu\nu}= {1\over 2} \Delta^{\mu\alpha} \Delta^{\nu\beta}
\left(\nabla_\alpha u_\beta - \nabla_\beta u_\alpha\right)\quad,
\ee
with
\be
A^{\langle\mu\nu\rangle} \equiv {1\over 2} \Delta^{\mu\alpha}\Delta^{\nu\beta}
\left(A_{\alpha\beta}+A_{\beta\alpha}\right) -{1\over 3} \Delta^{\mu\nu}\Delta^{\alpha\beta}
A_{\alpha\beta}\quad,
\ee
for any two-tensor $A^{\mu\nu}$.
The first order coefficient $\eta$ is the shear viscosity, while $\tau_{II}$, $\kappa$, and $\lambda_{1,2,3}$ are second order transport coefficients. As our 4D metric is flat and
the vorticity $\Omega^{\mu\nu}$ vanishes for our boost invariant plasma even with parity-asymmetry,
we won't have an access to $\kappa$ and $\lambda_{2,3}$ by our present expanding plasma.

The quest is to choose the right $\epsilon(\tau)$ and $u^\mu(\tau)$ as well as $\eta$, $\tau_{II}$,
and $\lambda_1$ to reproduce the energy-momentum tensor from the gravity side we obtained in the previous section. Since the gravity result already solves $D_\mu T^{\mu\nu}=0$, this automatically
includes the conservation equation in the hydrodynamics side too.
In fact, we would need one more piece of information; a general expectation is that local thermodynamics is completely specified by $\epsilon$ only\footnote{or equivalently by the local temperature $T$
which is related as $\epsilon = {3\pi^2 \over 8} N_c^2 T^4$ for ${\cal N}=4$ SYM.},
so that the transport coefficients should also be determined completely in terms of $\epsilon$ only. By
dimensional counting, one has in general conformal plasma,
\be
\eta=\eta^0 \epsilon^{3\over 4}\quad,\quad \lambda_1=\lambda_1^0 \epsilon^{1\over 2}
\quad,\quad \tau_{II}=\tau_{II}^0 \epsilon^{-{1\over 4}}\quad,\label{scaling}
\ee
with {\it fixed} dimensionless numbers $\eta^0$, $\lambda_1^0$, and $\tau_{II}^0$, which
are intrinsic to the microscopic details of the theory\footnote{These definitions are different
from those in Ref.\cite{Baier:2007ix}.}. At the end, the aim becomes : choose $\epsilon(\tau)$ and $u^\mu(\tau)$
as well as fundamental dimensionless constants $\eta^0$, $\lambda_1^0$, $\tau_{II}^0$
to match the energy-momentum tensor from the gravity result. The non-trivial test of the formalism
would be that $\eta^0$, $\lambda_1^0$, and $\tau_{II}^0$ should be same to those obtained from other methods/backgrounds.

From the discussion in section 2, we know that $\epsilon(\tau)$ and $u^y(\tau)$ start their expansion
as
\bear
\epsilon(\tau)&=& {3 w^4\over 16\pi G_5}\left({1\over \tau^{4\over 3}}+{\epsilon_1\over\tau^2}
+{\epsilon_2\over \tau^{8\over 3}}+\cdots\right)\quad,\nonumber\\
u^y(\tau)&=& -{C\over w^4 \tau^{5\over 3}}\left(1+{y_1\over\tau^{2\over 3}}+\cdots\right)+ {\cal O}(C^3)\quad,
\eear
where we restrict ourselves up to ${\cal O}(C^2)$ because our results in the previous section (\ref{result}) can tell things only up to this order. The $u^\tau$ is given from $u_\mu u^\mu =-1$ as
\be
u^\tau(\tau) = \sqrt{1+\tau^2\left(u^y(\tau)\right)^2}\quad.
\ee
It is computationally straightforward to insert the above general expansions into (\ref{proposal})
to have an energy-momentum tensor from SCVH, and to compare it with the gravity result (\ref{grav1}),
(\ref{result}) order by order
to see whether it works.

From comparing $T_{\tau\tau}$, one has
\bear
3w\epsilon_1+2 &=& 0 \quad \quad (\tau^{-2})\quad,\\
w^6\left(18 w^2 \epsilon_2 -1-2\log 2 \right) +12 C^2 &=& 0 \quad\quad (\tau^{-{8\over 3}})\quad,
\eear
and from $T_{\tau y}$,
\bear
y_1+\epsilon_1-\left({16\pi G_5\over 3}\right)^{1\over 4}{\eta^0\over w}&=& 0 \quad\quad (\tau^{-{5\over 3}})\quad,
\eear
where one can check that the next order $\tau^{-{7\over 3}}$ involves ${\cal O}(C^3)$ effects for which we
can't say much. From the comparison of $T_{yy}$, we have
\bear
2+w\epsilon_1 -4 \left({16\pi G_5\over 3}\right)^{1\over 4}\eta^0&=& 0 \quad\quad (\tau^0)\quad,\\
18w^4 \epsilon_2-5 w^2\left(1+2\log 2\right)-54\left({16\pi G_5\over 3}\right)^{1\over 4}w^3
\epsilon_1\eta^0 &&\nonumber \\
+48\left({16\pi G_5\over 3}\right)^{1\over 2}w^2\left(\lambda_1^0 -\eta^0\tau_{II}^0 \right)+{12 C^2\over w^4}&=& 0 \quad\quad (\tau^{-{2\over 3}})\quad,
\eear
and finally from $T_{ii}$ ($i=1,2$),
\bear
w\epsilon_1 +2\left({16\pi G_5\over 3}\right)^{1\over 4}\eta^0 &=& 0\quad\quad(\tau^{-2})\quad,\\
18w^4 \epsilon_2+w^2\left(1+2\log 2\right)+27\left({16\pi G_5\over 3}\right)^{1\over 4}w^3
\epsilon_1\eta^0 &&\nonumber \\
-24\left({16\pi G_5\over 3}\right)^{1\over 2}w^2\left(\lambda_1^0 -\eta^0\tau_{II}^0 \right)+{12 C^2\over w^4}&=& 0 \quad\quad (\tau^{-{8\over 3}})\quad.
\eear
One has five unknowns $\epsilon_{1,2}$, $y_1$, $\eta^0$, $\left(\lambda^0-\eta^0\tau_{II}^0\right)$
with the above seven equations to solve, and it is not a trivial thing for the SCVH to work, but
one indeed finds the consistent solution to the above,
\bear
\epsilon_1= -{2\over 3w}\quad,\quad \epsilon_2={(1+2\log 2)w^6 -12 C^2\over 18  w^8}\quad,
\quad y_1= {1\over w} \quad,\nonumber\\
 \eta^0={1\over 3}\left(3\over 16\pi G_5\right)^{1\over 4}\quad,\quad
\left(\lambda^0-\eta^0 \tau_{II}^0\right) = {\left(\log 2-1\right)\over 6}\left(3\over 16\pi G_5\right)^{1\over 2}\quad.\label{transsol}
\eear
To see the above results are in agreement with the literature, it is convenient to
rewrite things in terms of local temperature $T$ by using
\be
\epsilon = \left({3\over 16\pi G_5}\right)\pi^4 T^4\quad,
\ee
so that (\ref{scaling}) and (\ref{transsol}) give us
\be
\eta={\pi^2 T^3\over 16 G_5}\quad,\quad \left(\lambda_1 -\eta \tau_{II}\right)
={\left(\log 2-1\right)\over 2} {\pi T^2\over 16 G_5}\quad.\label{finsol}
\ee
Recalling that the area element at the horizon of a static black-hole with temperature $T$ is
\be
r_H^3 = \pi^3 T^3\quad,
\ee
so that the entropy density is related to the shear viscosity as
\be
s={r_H^3\over 4G_5} = {\pi^3 T^3 \over 4G_5} =(4\pi)\cdot\eta\quad,
\ee
which is a famous result.
The $\lambda_1$ and $\tau_{II}$ that are obtained in Ref.\cite{Baier:2007ix} for ${\cal N}=4$ SYM are
\be
\lambda_1= {\eta\over 2\pi T}\quad,\quad \tau_{II} = {\left(2-\log 2\right)\over 2\pi T}\quad,
\ee
which are again in agreement with (\ref{finsol}). Our main point is that the mapping works fine
even after our parity-asymmetric components included.

\section{Discussion}

We present a simple extension of previously studied boost invariant plasma by allowing parity asymmetric
components, and obtain its corresponding late time gravity solution up to second order expansion in the context
of AdS/CFT correspondence. We check that the second order conformal viscous hydrodynamics in
Ref.\cite{Baier:2007ix} consistently describes this plasma. There are a few directions that may be worth of
pursuing further; we haven't looked at the location of apparent horizon and the entropy density to see whether
the claim in Ref.\cite{Figueras:2009iu} is valid in our new background. Another direction would be to relax
assumptions in the gravity solution further to find more general gravity solutions, for example including radial
profile and/or anisotropicality in the transverse plane\footnote{See
Ref.\cite{Sin:2006pv,Janik:2008tc,Gubser:2009sx,Lin:2009pn} for some studies on anisotropic situations, and Ref.\cite{Pu:2008jf} for including radial profile.}. Numeric analysis
for early time dynamics within these more general setting in AdS/CFT may also be pursued further.

\vskip 1cm \centerline{\large \bf Acknowledgement} \vskip 0.5cm
We thank Sahoo Bindusar, Edi Gava, Kumar S. Narain, Dominik Nickel, M. M. Sheikh-Jabbari, Shigeki Sugimoto, and Mahdi Torabian for discussions.

\appendix
\section{Appendix}

In the appendix, we list the necessary expansion of the einstein equation up to second order,
after fixing the zero'th order solution
\be
a_0(u)=1-{w^4\over u^4}\quad,\quad b_0(u)=c_0(u)=F_0(u)=h_0(u)=0\quad.
\ee
The prime denotes derivative with respect to $u$.
\bear
E^\tau_\tau &=&\frac{2 w^4-8 u^5 a_1-7 u^6 a_1'-2 u^6 b_1'-2 u^2 w^4 b_1'-u^7 a_1''}{2 u^5 \tau^{2/3}}\nonumber\\
&+&\frac{-8 u^6 a_2-7 u^7 a_2'-2 u^7 b_2'-2 u^3 w^4 b_2'-u^8 a_2''}{2 u^6 \tau^{4/3}}
+\frac{1}{6 u^8 \tau^{4/3}}\Bigg(-6 u^2 w^4+6 u^7 a_1\nonumber\\&+&4 u^7 b_1-24 u^8 F_1^2-48 u^8 F_1h_1-24 u^8 h_1^2 +12 u^4 w^4 h_1^2-12 w^8 h_1^2+ 3 u^8 a_1'-2 u^8 b_1'\nonumber\\&-&6 u^9a_1 b_1'-3 u^{10} a_1' b_1'+6 u^8 c_1'-27 u^9 F_1 F_1'-6 u^9 h_1 F_1'-6 u^5 w^4 h_1 F_1'-3 u^{10}F_1'^2\nonumber\\&-&6 u^9 F_1 h_1'
-6 u^5 w^4 F_1 h_1'-6 u^9 h_1 h_1'+6 u w^8 h_1 h_1'-3 u^{10}F_1 F_1''\Bigg)
\quad,\nonumber\\
E^\tau_u&=& \frac{-2 b_1'-u b_1''}{u \tau^{1/3}}+\frac{-2 b_2'-u b_2''}{u \tau}
+\frac{1}{2 u^6 \tau}\Bigg(-12 w^4 h_1^2+4 u^4 b_1'-2 u^6 b_1'^2-4 u^4 c_1'+4 u^6 b_1' c_1'\nonumber\\
&-&3 u^6 c_1'^2-u^5 h_1 F_1'-6 u^5 F_1 h_1'-6 u^5 h_1h_1'+6 u w^4 h_1 h_1'+u^6 h_1 F_1''\Bigg)\quad,\nonumber\\
E^u_\tau &=& \frac{-4 w^4+3 u^5 a_1-4 u w^4 b_1+3 u^6 c_1'-3 u^2 w^4 c_1'}{3 u^4 \tau^{5/3}}
+\frac{1}{3 u^5 \tau^{7/3}}\Bigg(6 u^6 a_2-8 u^2 w^4 b_2\nonumber\\&+&
2 u^7 b_2'-2 u^3 w^4 b_2'+3 u^7 c_2'
-3 u^3 w^4 c_2'\Bigg)
+\frac{1}{9 u^7 \tau^{7/3}}\Bigg(12 u^2 w^4-3 u^7 a_1-4 u^7 b_1\nonumber\\&+&6 u^3 w^4 b_1-6 u^7 c_1-6 u^3 w^4 c_1+18 u^8 F_1^2+36 u^8 F_1 h_1-36 u^4 w^4 F_1 h_1+18 u^8 h_1^2\nonumber\\&-&36 u^4 w^4 h_1^2+18 w^8 h_1^2-3 u^8 a_1'-3 u^9 b_1 a_1'+8 u^8 b_1'-6 u^4 w^4 b_1'+3 u^9 a_1 b_1'\nonumber\\&+&6 u^9 b_1 b_1'-6 u^5 w^4 b_1 b_1'-6 u^9 c_1 b_1'+6 u^5 w^4 c_1 b_1'+6 u^4 w^4 c_1'+9 u^9 a_1 c_1'-6 u^9 b_1 c_1'\nonumber\\&+&6 u^5 w^4 b_1 c_1'+9 u^9 c_1 c_1'-9 u^5 w^4 c_1 c_1'\Bigg)\quad,\nonumber\\
E^u_u&=&
\frac{1}{2 u^5 \tau^{2/3}}\left(2 w^4-8 u^5 a_1-7 u^6 a_1'-6 u^6 b_1'+2 u^2 w^4 b_1'-u^7 a_1''-2 u^7 b_1''+2 u^3 w^4 b_1''\right)\nonumber\\
&+&\frac{1}{2 u^6 \tau^{4/3}}\left(-8 u^6 a_2-7 u^7 a_2'-6 u^7 b_2'+2 u^3 w^4 b_2'-u^8 a_2''-2 u^8 b_2''+2 u^4 w^4 b_2''\right)\nonumber\\
&+&
\frac{1}{6 u^8 \tau^{4/3}}\Bigg(-6 u^2 w^4+6 u^7 a_1+4 u^7 b_1-24 u^8 F_1^2-48 u^8 F_1 h_1-24 u^8 h_1^2
\nonumber\\&-&24 u^4 w^4 h_1^2+24 w^8 h_1^2+3 u^8 a_1'+2 u^8 b_1'-12 u^4 w^4 b_1'-18 u^9 a_1 b_1'-3 u^{10} a_1' b_1'-6 u^{10} b_1'^2\nonumber\\&+&6 u^6 w^4 b_1'^2-6 u^8 c_1'+12 u^4 w^4 c_1'+12 u^{10} b_1' c_1'-12 u^6 w^4 b_1' c_1'-9 u^{10} c_1'^2+9 u^6 w^4 c_1'^2\nonumber\\&-&42 u^9 F_1 F_1'-24 u^9 h_1 F_1'+12 u^5 w^4 h_1 F_1'-3 u^{10} F_1'^2-24 u^9 F_1 h_1'+12 u^5 w^4 F_1 h_1'\nonumber\\&-&24 u^9 h_1 h_1'+36 u^5 w^4 h_1 h_1'-12 u w^8 h_1 h_1'-4 u^9 b_1''-6 u^{10} a_1 b_1''-6 u^{10} F_1 F_1''\Bigg)\,,\nonumber\\
E^y_y&=&
\frac{1}{u^3 \tau^{2/3}}\left(-4 u^3 a_1-u^4 a_1'-6 u^4 b_1'+2 w^4 b_1'+5 u^4 c_1'-w^4 c_1'-u^5 b_1''+u w^4 b_1''+u^5 c_1''-u w^4 c_1''\right)\nonumber\\
&+&\frac{1}{u^3 \tau^{4/3}}\left(-4 u^3 a_2-u^4 a_2'-6 u^4 b_2'+2 w^4 b_2'+5 u^4 c_2'-w^4 c_2'-u^5 b_2''+u w^4 b_2''+u^5 c_2''-u w^4 c_2''\right)\nonumber\\
&+&\frac{1}{6 u^8 \tau^{4/3}}\Bigg(24 u^7 a_1+16 u^7 b_1-12 u^7 c_1-24 u^8 F_1^2-48 u^8 F_1 h_1-24 u^8 h_1^2+12 u^4 w^4 h_1^2\nonumber\\&+&12 w^8 h_1^2+6 u^8 a_1'-4 u^8 b_1'-12 u^4 w^4 b_1'-36 u^9 a_1 b_1'-6 u^{10} a_1' b_1'-6 u^{10} b_1'^2+6 u^6 w^4 b_1'^2\nonumber\\&+&2 u^8 c_1'+6 u^4 w^4 c_1'+30 u^9 a_1 c_1'+6 u^{10} a_1' c_1'+6 u^{10} b_1' c_1'-6 u^6 w^4 b_1' c_1'-27 u^9 F_1 F_1'\nonumber\\&-&6 u^9 h_1 F_1'+6 u^5 w^4 h_1 F_1'-3 u^{10} F_1'^2-6 u^9 F_1 h_1'+6 u^5 w^4 F_1 h_1'-6 u^9 h_1 h_1'+12 u^5 w^4 h_1 h_1'\nonumber\\&-&6 u w^8 h_1 h_1'-4 u^9 b_1''-6 u^{10} a_1 b_1''+4 u^9 c_1''+6 u^{10} a_1 c_1''-3 u^{10} F_1 F_1''\Bigg)\quad,\nonumber\\
E^i_i&=&
\frac{1}{2 u^5 \tau^{2/3}}\left(-2 w^4-8 u^5 a_1-2 u^6 a_1'-2 u^6 b_1'+2 u^2 w^4 b_1'-5 u^6 c_1'+u^2 w^4 c_1'-u^7 c_1''+u^3 w^4 c_1''\right)\nonumber\\
&+&\frac{1}{2 u^6 \tau^{4/3}}\left(-8 u^6 a_2-2 u^7 a_2'-2 u^7 b_2'+2 u^3 w^4 b_2'-5 u^7 c_2'+u^3 w^4 c_2'-u^8 c_2''+u^4 w^4 c_2''\right)\nonumber\\
&+&\frac{1}{6 u^8 \tau^{4/3}}\Bigg(6 u^2 w^4+6 u^7 a_1+4 u^7 b_1+6 u^7 c_1-24 u^8 F_1^2-48 u^8 F_1 h_1-24 u^8 h_1^2\nonumber\\&+&12 u^4 w^4 h_1^2+12 w^8 h_1^2-2 u^8 b_1'-6 u^9 a_1 b_1'-u^8 c_1'-3 u^4 w^4 c_1'-15 u^9 a_1 c_1'-3 u^{10} a_1' c_1'\nonumber\\&-&3 u^{10} b_1' c_1'+3 u^6 w^4 b_1' c_1'-12 u^9 F_1 F_1'-6 u^9 h_1 F_1'+6 u^5 w^4 h_1 F_1'-6 u^9 F_1 h_1'+6 u^5 w^4 F_1 h_1'\nonumber\\&-&6 u^9 h_1 h_1'+12 u^5 w^4 h_1 h_1'-6 u w^8 h_1 h_1'-2 u^9 c_1''-3 u^{10} a_1 c_1''\Bigg)\quad,\nonumber\\
E^\tau_y &=&
\frac{1}{2} \left(5 u F_1'+u^2 F_1''\right) \tau^{1/3}+\frac{1}{6 u \tau^{1/3}}\Bigg(9 F_1+12 h_1-30 u^2 F_1 b_1'+30 u^2 F_1 c_1'+12 u F_1'\nonumber\\&-&3 u^3 b_1' F_1'+6 u^3 c_1' F_1'+15 u^2 F_2'+4 u h_1'-6 u^3 F_1 b_1''+6 u^3 F_1 c_1''+3 u^2 F_1''+3 u^3 F_2''\Bigg)\quad,\nonumber\\
E^u_y&=& \frac{1}{9 \tau^{4/3}}\left(3 F_1+4 h_1-6 u^2 F_1 b_1'+6 u^2 F_1 c_1'+6 u F_1'+3 u^2 b_1 F_1'+3 u^2 F_2'\right)\quad,\nonumber\\
E^y_\tau&=&
\frac{-5 u^4 F_1'+5 w^4 F_1'-u^5 F_1''+u w^4 F_1''}{2 u^3 \tau^{5/3}}
+\frac{1}{6 u^5 \tau^{7/3}}\Bigg(-9 u^4 F_1+3 w^4 F_1-12 u^4 h_1+12 w^4 h_1\nonumber\\&+&15 u^6 F_1 a_1'-12 u^2 w^4 F_1 b_1'+24 u^2 w^4 F_1 c_1'+8 u^5 F_1'-18 u w^4 F_1'-15 u^6 a_1 F_1'+30 u^6 b_1 F_1'\nonumber\\&-&30 u^2 w^4 b_1 F_1'-30 u^6 c_1 F_1'+30 u^2 w^4 c_1 F_1'+3 u^7 b_1' F_1'-3 u^3 w^4 b_1' F_1'-6 u^7 c_1' F_1'+6 u^3 w^4 c_1' F_1'\nonumber\\&-&15 u^6 F_2'+15 u^2 w^4 F_2'-4 u^5 h_1'+4 u w^4 h_1'+3 u^7 F_1 a_1''+u^6 F_1''-3 u^2 w^4 F_1''-3 u^7 a_1 F_1''\nonumber\\&+&6 u^7 b_1 F_1''-6 u^3 w^4 b_1 F_1''-6 u^7 c_1 F_1''+6 u^3 w^4 c_1 F_1''-3 u^7 F_2''+3 u^3 w^4 F_2''\Bigg)\,,\nonumber\\
E^y_u&=&
\frac{5 F_1'+u F_1''}{2 u \tau^{4/3}}+\frac{1}{6 u^7 \tau^2}\Bigg(9 u^4 F_1+12 u^4 h_1-6 w^4 h_1+15 u^6 h_1 a_1'-18 u^6 F_1 b_1'-18 u^6 h_1 b_1'\nonumber\\&+&6 u^2 w^4 h_1 b_1'+30 u^6 F_1 c_1'+30 u^6 h_1 c_1'-6 u^2 w^4 h_1 c_1'-18 u^5 F_1'-30 u^6 b_1 F_1'+30 u^6 c_1 F_1'\nonumber\\&-&3 u^7 b_1' F_1'+6 u^7 c_1' F_1'+15 u^6 F_2'+4 u^5 h_1'+3 u^7 h_1 a_1''+6 u^7 F_1 c_1''+6 u^7 h_1 c_1''-6 u^3 w^4 h_1 c_1''\nonumber\\&-&3 u^6 F_1''-6 u^7 b_1 F_1''+6 u^7 c_1 F_1''+3 u^7 F_2''\Bigg)\quad.\nonumber
\eear

 \vfil

\end{document}